\documentclass[pra,aps,showpacs,twocolumn]{revtex4}
\usepackage{epsfig}
\usepackage{bm}

\pacs{67.85.-d,03.75.Lm,03.75.Hh}  

\begin{document} 
\title{Two-dimensional dynamics of expansion of a degenerate Bose gas} 

\author{ Igor E. Mazets$^{1,2}$}
\affiliation{$^1$Vienna Center for Quantum Science and Technology, Atominstitut, TU Wien, 1020 Vienna, Austria \\
$^2$Ioffe Physico-Technical Institute of the Russian Academy of Sciences, 194021 St.Peterburg, Russia } 

\begin{abstract}
Expansion of a degenerate Bose gas released from a pancake-like trap is numerically simulated under assumption 
of separation of motion in the plane of  the loose initial trapping and the motion in the direction of the initial 
tight trapping. The initial conditions for the phase fluctuations are generated using the extension to the two-dimensional case 
of the description 
of the phase noise by the Ornstein-Uhlenbeck stochastic process. The numerical simulations, 
taking into account both the finite size of the two-dimensional system and the atomic interactions, which cannot be 
neglected on the early stage of expansion, did not reproduce the scaling law for the peaks in the 
density fluctuation spectra experimentally observed by Choi, Seo, Kwon, and Shin [Phys. Rev. Lett. \textbf{109}, 125301 (2012)]. 
The latter experimental results   may thus require an explanation beyond our current assumptions.  
\end{abstract} 

\maketitle 

Correlations in degenerate Bose gases with repulsive interactions manifest themselves most apparently via phase coherence. Even 
low-dimensional degenerate bosonic systems (quasicondensates) demonstrate phase coherence over finite distances. 
After release from the trap, phase fluctuations are converted, in the course of free expansion, into density-density correlations 
(density ripples). Theory for expansion of both one-dimensional (1D) and two-dimensional (2D) trapping geometries has been 
developed \cite{Manz1} and experimentally proven for 1D trapped gases \cite{Manz2} 
(similar behavior has been previously found in expanding clouds, 
which were in the three-dimensional Bose-Einstein condensate rather in the 1D regime inside a very elongated trap \cite{Hanno}).

In contrast to 1D Bose gases, where correlations decay 
exponentially and their characteristic length $\lambda _T$ determines  
the typical wavelength scale of the density ripples and the timescale $m\lambda _T^2/\hbar $ of their emergence ($m$ being the mass 
of the atom), correlations in degenerate 2D Bose gases decay with the distance according to a power law, if the temperature is 
below the point of the Beresinskii-Kosterlitz-Thouless (BKT) transition \cite{BKT-s}. Such a power-law decay renders no specific 
correlation length, the spectrum of the density fluctuations of a 2D gas released from a trap evolves in a self-similar way 
at asymptotically large expansion times $t_\mathrm{e}$. In particular, after averaging the power spectrum of density 
fluctuations 
\begin{equation} 
P_\mathbf{q} =\left| \int d^2 \bm{\rho }\,  \delta n(\bm{\rho }) e^{-i\mathbf{q} \bm{\rho }}\right| ^2 ,
\label{Pq} 
\end{equation} 
where $\delta n(\bm{\rho })$ is the local density fluctuation, over the directions of the 2D wave vector \textbf{q}, 
the $n$th peak position $q_n$, $n=1,2,3,\, \dots \, $, is expected to satisfy the relation \cite{Manz1} 
\begin{equation} 
\hbar q_n^2t_\mathrm{e}/(2\pi m) = n-1/2 . 
\label{qn-id} 
\end{equation} 

Recently, Choi \textit{et al.} \cite{Choi} reported their experimental results on probing fluctuations in a 2D gas of bosonic 
($^{23}$Na) atoms in free expansion after a sudden release from a pancake-like optical trap, the trapping 
frequencies being equal to $(\omega_x,\, \omega_y,\, \omega _z)=2\pi \times (3.0,\, 3.9,\, 370)$~Hz. The chemical potential 
was about $\mu _0=2\pi \hbar \times 260$~Hz, i.e., less than the spacing between the discrete levels of the trapping 
Hamiltonian in the tighgtly confined direction. 
The coherent fraction of atomic ensemble decreased from 0.78 to 0.12 as the temperaure $T$ grew  from 20~nK up to 
the BKT transition temperature $T_\mathrm{BKT}\approx 70$~nK. 

The experimental results \cite{Choi} turned out to be quite surprising. Instead of recovering the scaling 
(\ref{qn-id}), they brought about another dependence, 
\begin{equation} 
\hbar q_n^2t_\mathrm{e}/(2\pi m)=\alpha n^\gamma  ,  
\label{qn-exp} 
\end{equation} 
with $0.2<\alpha <0.45$ and $0.7<\gamma <1$ for expansion times in the range $10~\mathrm{ms}<t_\mathrm{e}<25$~ms. 
In other words, the spectral peaks of the density fluctuations were not only shifted with respect to Eq.~(\ref{qn-id}), 
but the spacing between the ajacent peaks was significantly smaller than Eq. (\ref{qn-id}) predicts. 
Choi \textit{et al.} ruled out the effects of the cloud finite size 
and suggested that such an unexpected behavior might be intrinsic to the expansion dynamics \cite{Choi}. 

To check this assumption, we performed numerical simulations of 2D expansion of a degenerate gas. 
Since the temperature and, hence, the coherent fraction do not 
affect significantly the peak positions $q_n$ \cite{Choi}, we find it safe to neglect thermal population 
of the excited levels of motion in the tightly trapped ($z$) direction and model the system's dynamics by 
the Gross-Pitaevskii equation (GPE), 
taking into account the interaction effects, which are important at the initial stage of expansion. 
Furthemore, we separate the motion in $z$-direction and in the $(x,y)$-plane. This would be impossible if the Beliaev and Landau 
damping of elementary excitations in expanding but still dense enough atomic cloud were giving rise to significant population  
of modes with non-zero $z$-components of kinetic momentum. However, the estimation of the ratio of the damping 
rate of an elementary excitation to its frequency yields \cite{damp} the value 
$\sim a_s^{1/2}/(n_\mathrm{3D}^{1/2}\lambda _\mathrm{th}^2)\sim 10^{-2}$, where $a_s$ is the atomic $s$-wave scattering length, 
$n_\mathrm{3D}$ is the three-dimensional atomic number density, and $\lambda _\mathrm{th}$ is the de Broglie wave length of 
atoms at temperature $T$. Even for elementary excitations with the energy close to the chemical potential the typical 
relaxation time (and, hence, the characteristic time of scattering into $z$-direction) in a three-dimensional degenerate gas 
with the bulk density equal to the peak density in our 2D system is of the order of 100~ms, which is much 
longer than the typical expansion time. Therefore we can separate the motion in different directions. In $z$ direction, we have a 
free expansion of a  wave packet, which is initially, at $t=0$, the wave function of the ground state of the harmonic 
trapping potential with the fundamental frequency $\omega _z$. Then the motion in the $(x,y)$ plane is described by the GPE 
\begin{equation} 
i\hbar \frac {\partial }{\partial t} \Psi =-\frac {\hbar ^2}{2m} \left( \frac {\partial ^2}{\partial x^2}+
\frac {\partial ^2}{\partial y^2}\right) \Psi +\mu (t) |\Psi |^2 \Psi . 
\label{GPE2D} 
\end{equation} 
The fast (but still not infinitely fast) evolution of the quasicondensate profile in $z$-direction  affects the slow motion in the perpendicular plane) only through the time-dependent nonlinearity in Eq. (\ref{GPE2D}).  
The time-dependent effective 2D coupling constant $\mu (t)$ takes into account the growth of the transverse density profile 
width $\propto \sqrt{1+(\omega _zt)^2}$. 
The normalization condition $\mathrm{max}\, |\Psi (x,y,{t=0})| = |\Psi (x=0,y=0,{t=0})|\equiv 1$ fixes the 
prefactor, and we obtain 
\begin{equation} 
\mu (t) =\frac {\mu _0}{\sqrt{ 1+(\omega _zt)^2}} . 
\label{coupling-t} 
\end{equation} 

Initial shape of $|\Psi |^2$ is the inverted-parabolic Thomas-Fermi profile corresponding to the chemical potential $\mu _0$ 
of the trapped gas of sodium atoms. The phase $\varphi =\arg \Psi $ at $t=0$ can be generated using the generalization 
of the stochastic Ornstein-Uhlenbeck process, previously used to model random phase distributions of phases in 1D quasicondensates 
\cite{OU1,OU2}. 

Assume that the temperature $T$ of 2D Bose gas is below the BKT transition,  phase and density fluctuations are small and, 
therefore, the Hamiltonian can be linearized: 
\begin{equation} 
H= \frac {\hbar ^2}{2m} \int d^2 \bm{\rho } \left[ n_\mathrm{2D}(\bm \rho ) (\nabla \! _\perp \varphi )^2 +\tilde{g}\,  
\delta n^2+\frac { (\nabla \! _\perp \, \delta n )^2}{4n_\mathrm{2D}(\bm \rho )}\right] .
\label{Hlin}
\end{equation}  
Here $n_\mathrm{2D} =\int dz\, n_\mathrm{3D}$ is the 2D number density, $\tilde{g}$ is the dimensionless coupling strength 
($\tilde{g}\approx 0.013$ in the experiment \cite{Choi}), $\nabla \! _\perp $ is the 2D gradient 
in the $(x,y)$-plane. Then the partition function of a trapped gas is 
\begin{equation} 
Z =\int {\cal D}\varphi \int {\cal D} \delta n \, \exp [ -H/(k_\mathrm{B}T)] . 
\label{Z1} 
\end{equation} 
The density fluctuation variable can be integrated out, and we obtain 
\begin{equation} 
Z =\mathrm{const} \int {\cal D}\varphi  \, \exp \left[ -\frac {\hbar ^2}{2mk_\mathrm{B}T}\int d^2 \bm{\rho } \, 
n_\mathrm{2D}(\bm \rho ) (\nabla \! _\perp \varphi )^2  \right] . 
\label{Z2} 
\end{equation} 

We can use the thermodynamic expression (\ref{Z2}) to generate initial phase distribution for individual realizations 
of our numerical simulations. On a $(2M+1)\times (2M+1)$ square grid Eq. (\ref{Z2}) takes the form 
\begin{eqnarray} 
Z &=&\mathrm{const} \prod _{l_x=-M}^M\prod _{l_y=-M}^M \int d\varphi _{l_x\, l_y} \, \exp \Bigg {\{ } -\frac \varepsilon 2 
w _{l_x\, l_y} \times \nonumber \\
&& [(\varphi _{l_x\, l_y}-\varphi _{l_x\, l_y-1})^2+(\varphi _{l_x\, l_y}-\varphi _{l_x-1\, l_y})^2]\Bigg {\} } .
\label{grid} 
\end{eqnarray} 
Here the pair of subscripts $l_x$ and $l_y$ denotes a variable at the point with coordinates $x=l_x\, \Delta s$ and 
$y=l_y\, \Delta s$ [note that the grid step $\Delta s$ does not appear in Eq. (\ref{grid}) in our 2D problem]. 
The density profile and its peak value are characterized by dimensionless quantities 
$w _{l_x\, l_y} =n_\mathrm{2D} (l_x\, \Delta s,\, l_y\, \Delta s)/
n_\mathrm{2D} (0,0)$ and $\varepsilon = \hbar ^2 n_\mathrm{2D} (0,0)/(mk_\mathrm{B}T)$, respectively. 
Periodic boundary conditions are set, i.e., 
$\varphi _{l_x\, M+1}=\varphi _{l_x\, -M}$ and $\varphi _{M+1\, l_y}=\varphi _{-M\, l_y}$.

Assume we know the $2M+1$ phases $\varphi _{l_x-1\, l_y}$. 
Then we can calculate the phases $\varphi _{l_x\, l_y}$ in the next row of the grid using the transfer matrix 
approach \cite{transfm}. 

We assume that the grid step is small enough and thus we can write $w _{l_x\, l_y}/w _{l_x\, 0}\approx w _{l_x-1\, l_y}/w _{l_x-1\, 0}$. 
The transformation \cite{prim1} 
\begin{equation} 
A_{l_x-1\, j}=\sum _{l_y =-M}^M V_{l_x; j\, l_y} \sqrt{\frac {w _{l_x\, l_y}}{w _{l_x\, 0}}}\varphi _{l_x-1\, l_y}   
\label{td} 
\end{equation} 
diagonalizes the non-negative quadratic form 
\begin{equation} 
\sum _{l_y=-M}^M\frac {w _{l_x\, l_y}}{w _{l_x\, 0}}(\varphi _{l_x-1\, l_y}-\varphi _{l_x-1\, l_y-1})^2=
\sum _{j=0}^{2M}J_{l_x\, j} A_{l_x-1\, j }^2 . 
\label{q-form}
\end{equation} 
Both $V_{l_x; j\, l_y} $ and $J_{l_x\, j}$ can found numerically by standard methods of linear algebra. Then Eq. (\ref{grid}) factorizes:   
$Z=\mathrm{const} \prod _{j=0}^{2M} Z_j$, where each factor 
\begin{eqnarray} 
Z_j&=&\prod _{l_x=-M}^M \int dA_{l_x\, j}\, \exp \Bigg {[} -\frac {\varepsilon _{l_x}}2 ( A_{l_x-1\, j}-A_{l_x\, j})^2-\nonumber \\
&& \frac {\varepsilon _{l_x}J_{l_x\, j}}2A_{l_x\, j}^2\Bigg {]} 
\label{Zj} 
\end{eqnarray} 
corresponds to a 1D evolution (along the $x$-coordinate) of a $A_j$ describable by the Ornstein-Uhlenbeck stochastic process (which can be easily seen 
from the analogy with the relative phase between two tunnel-coupled 1D quasicondensates at thermal equilibrium \cite{OU1}). Here $\varepsilon _{l_x} = 
\varepsilon w_{l_x\, 0}$. The formula \cite{G96} 
\begin{equation} 
A_{l_x\, j}=A_{l_x-1\, j}e^{-\sqrt{J_{l_x\, j}}} +\sqrt{ \frac {1-e^{-2\sqrt{J_{l_x\, j}}}}{ 2\varepsilon _{l_x}\sqrt{J_{l_x\, j}} } }\varsigma , 
\label{update}
\end{equation} 
where $\varsigma $ is a (pseudo)random number obeying the Gaussian distribution with zero mean and unity variance,  
updates the variable $A_j$ on the next step. Then we obtain the initial phase values in the next grid row: 
\begin{equation} 
\varphi _{l_x\, l_y} =\sum _{l_y =-M}^M V_{l_x; j\, l_y} \sqrt{\frac {w _{l_x\, 0}}{w _{l_x\, l_y}}} A_{l_x\, j} .
\label{obt} 
\end{equation} 
Subsequent steps are done using the new local values for the density profile. The same procedure can be used 
to obtain $A_{l_x-2\, j}$
from the known $A_{l_x-1\, j}$. 

Eq. (\ref{update}), together with chosing the (pseudo)random initial values of the $A_{l_x\, j}$ variables for a certain row 
(e.g., $l_x=0$) according  to a Gaussian distribution with zero 
mean and variance $\langle A_{l_x\, j}^2\rangle =1/(2\varepsilon _{l_x}\sqrt{J_{l_x\, j}})$, enables us to prepare the initial conditions for 
each realization of our numerical modeling \cite{prim2}. 

\begin{figure}[t] 

\centerline{\epsfig{file=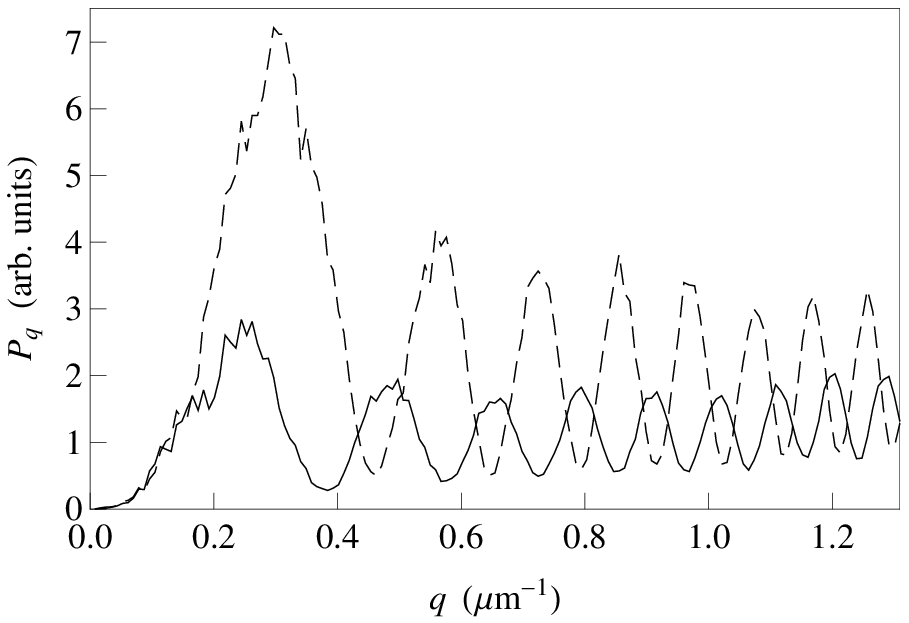,width=0.95\columnwidth} }

\begin{caption}
{\label{obr1} Solid line: the power spectrum of density fluctuations for $t_\mathrm{e}=11$~ms obtained from Eq. (\ref{GPE2D}), i.e., taking into account 
interaction of atoms in an expanding atomic cloud. Parameters of the trapped degenerate cloud are taken from Ref. \cite{Choi} (see details in the text). 
Dashed line: spectrum of density fluctuations obtained for a purely ballistic expansion during the same $t_\mathrm{e}$. The spectra are averaged over 5 
realisations each. }
\end{caption} 
\end{figure} 

\begin{figure}[b] 

\centerline{\epsfig{file=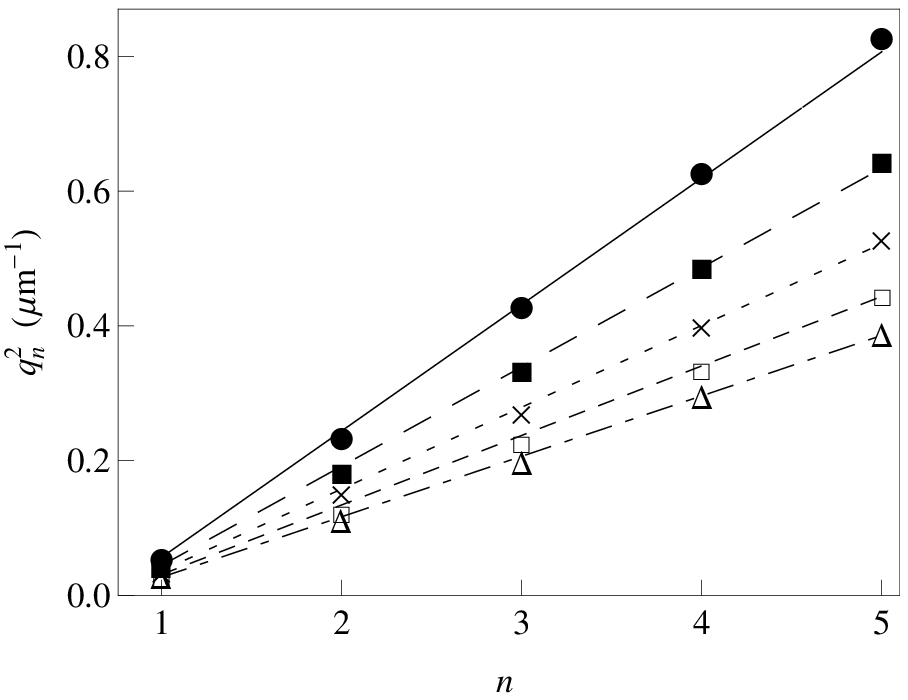,width=0.95\columnwidth} }

\begin{caption} 
{\label{obr2} The peak positions for the power spectrum of density fluctuations (the peak number $n$ is dimensionless) 
of a $^{23}$Na ultracold gas interacting during the expansion for 
$t_\mathrm{e}=$ 11 ms (filled circles), 14 ms (filled squares), 17 ms (crosses), 20 ms (open squares), and 23 ms (triangles). Straight lines 
(solid, long-dashed, dotted, short-dashed, and dot-dashed, respectively) display the fitting by Eqs. (\ref{qn-calc},~\ref{bel}). }
\end{caption} 
\end{figure} 

We solved Eq. (\ref{GPE2D}) using the  split-step pseudospectral method for the typical parameters of Ref. \cite{Choi} and obtained the power spectra 
of density fluctuations. As the density fluctuation $\delta n$ 
in Eq. (\ref{Pq}) we took the difference between the local densities given at the same expansion time $t_\mathrm{e}$  by two solutions, one containing initial 
phase fluctuations and other having constant phase everywhere at $t=0$. An example of such a spectrum $P_q$, that is $P_\mathbf{q}$ averaged over the direction 
of the wave vector \textbf{q}, is displayed in Fig.~\ref{obr1}. The variation of the temperature from 60~nK to 20~nK (corresponding to the increase of 
$\varepsilon $ from 15 to 45) changes the height of the peaks of $P_q$ only, leaving the peak positions $q_n$ unmodified, 
as we can expect from the experimental results \cite{Choi}.  
The solution of the GPE (\ref{GPE2D}) is also juxtaposed with the density fluctuations spectra for purely ballistic expansion [i.e., by setting 
$\mu (t)\equiv 0$]. In the case of purely ballistic expansion, we found that Eq. (\ref{qn-id}) is satisfied with 1\% accuracy for $n\geq 2$; for the first peak 
position we obtained $\hbar q_1^2 t_\mathrm{e}/(2\pi m)\approx 0.4$ for  $10~\mathrm{ms}\leq t_\mathrm{e}\leq 23$~ms.

As we can see from Fig.~\ref{obr1}, interaction of atoms during expansion is not negligible and shifts the peak positions with respect to the 
ballistic expansion case. However, instead of recovering Eq. (\ref{qn-exp}), we found that the peak positions are well described by the formula 
\begin{equation} 
\hbar q_n^2 t_\mathrm{e}/(2\pi m)=\beta (n-\ell ) , 
\label{qn-calc}
\end{equation} 
where 
\begin{equation} 
\beta \approx 0.9,\qquad \ell \approx 0.7
\label{bel} 
\end{equation} 
for the whole range of experimentally relevant expansion times (see Fig.~\ref{obr2}). 

Note, that the transversal trapping frequency $\omega _z=2\pi \times 370$~Hz of Ref. \cite{Choi} is approximately equal to the 
radial trapping frequency in the experiment of Ref. \cite{Hanno}, but the cigar-shaped configuration of the atomic cloud resulted 
in the latter case in a faster 
($\propto t^{-2}$ instead of $\propto t^{-1}$ at  expansion times $\gtrsim 1$~ms) decrease of the three-dimensional 
density of atoms than in the 2D case. Radial trapping frequencies in recent experiments with 1D ultracold gases on atom chips \cite{Manz2,OU2} were by an 
order of magnitude higher, thus further decreasing the time scale, on which atomic interactions still play a role. Therefore expansion of atomic clouds in 
the experiments \cite{Manz2,Hanno,OU2} was very close to ballistic, in contrast to the case modeled in our present work.

Our numerical results do not reproduce the observations by Choi \textit{ et al.} \cite{Choi}. 
For example, our calculations predict the values of the fifth peak position $q_5$, 
which are at relevant expansion times larger by a factor 2--3 than the observed ones. In general, our calculations predict the linear depenfdence of $q_n^2$ on the 
peak number $n$, in contrast to the nonlinear fitting formula (\ref{qn-exp}) found in Ref. \cite{Choi}. 

Such a conspicuous disagreement between the theory that assumes separation of motion in the former trap plane and in the direction of the tight confinement before 
the release of the atom from the trap and the experimental results \cite{Choi} calls for explanantion. We do not expect the scattering of phonons into 
$z$-direction in the expanding quasicondensate to be a reason for such a discrepancy, since the estimations based on 
the theory of Ref. \cite{damp} show that this process is too 
slow compared to the typical expansion times. Moreover, the influence of such a scattering would grow with $t_\mathrm{e}$, but our results deviate strongly from the 
experimental findins at shorter times as strongly as at longer ones. 

The fact that the experimental spectra of density fluctuations are describable by Eq. (\ref{qn-exp}) when no vortices are present in the 2D system \cite{Choi} allow 
us to rule out any vortex-related explanation of this discrepancy, which therefore still remains puzzling. 

This work was supported by the FWF (project P~22590-N16). The author thanks J. Schmiedmayer and Yong-il Shin 
for helpful discussions.


\begin{thebibliography}{99}
\bibitem{prim2} As a consistency check of our method, we can demonstrate that the description via the Ornstein-Uhlenbeck stochastic process yields the 
correct correlation function of a uniform 2D gas  
$\langle \exp [ i\varphi (0,0) -i\varphi (x,y) ] \rangle \propto \left( \sqrt{ x^2+y^2} \right) ^{-1/(2\pi \varepsilon )}$.
\end{thebibliography}
\end{document}